\def\mdot{$M_\odot \, {\rm yr}^{-1}$}
\providecommand{\adsurl}[1]{\href{#1}{ADS}}

\documentclass[iop,twocolumn,numberedappendix]{emulateapj} 
\usepackage{graphics,graphicx,epsf,natbib}
\usepackage{subfigure}
\usepackage[colorlinks=true,linkcolor=blue,anchorcolor=blue,citecolor=blue,urlcolor=blue]{hyperref}
\usepackage{color}
\usepackage{graphicx}
\usepackage{amssymb,amsmath}
\usepackage{natbib}

\slugcomment{Accepted ApJ}

\shorttitle{UV Morphologies and Star-Formation Rates of CLASH BCGs}
\shortauthors{Donahue et al.}

\begin{document}


\title{Ultraviolet Morphologies and Star-Formation Rates of CLASH Brightest Cluster Galaxies}


\author{Megan Donahue\altaffilmark{1},
				Thomas Connor\altaffilmark{1}, 
				Kevin Fogarty\altaffilmark{2},
				Yuan Li\altaffilmark{3},
              G. Mark Voit\altaffilmark{1}, 
                 Marc Postman\altaffilmark{2}, \\
               Anton Koekemoer\altaffilmark{2},
               John Moustakas\altaffilmark{4}, Larry Bradley\altaffilmark{2}, Holland Ford\altaffilmark{5}
              }

\altaffiltext{1}{Physics and Astronomy Dept., Michigan State University, East Lansing, MI, 48824 USA; \email{donahue@pa.msu.edu}}
\altaffiltext{2}{Space Telescope Science Institute, 3700 San Martin Drive, Baltimore, MD 21218 USA}
\altaffiltext{3}{Astronomy Dept., University of Michigan, Ann Arbor, MI}
\altaffiltext{4}{Dept of Physics and Astronomy, Siena College, Loudonville, NY USA}
\altaffiltext{5}{Dept of Physics and Astronomy, The Johns Hopkins University, Baltimore, MD 21218 USA}

\begin{abstract}

Brightest cluster galaxies (BCGs) are usually quiescent, but many exhibit star formation. 
Here we exploit the opportunity provided by rest-frame UV imaging of galaxy clusters in the CLASH (Cluster Lensing and Supernovae with Hubble) Multi-Cycle Treasury Project to reveal the diversity of UV morphologies in BCGs and to compare them with recent simulations of the cool, star-forming gas structures produced by precipitation-driven feedback.
All of the CLASH BCGs are detected in the rest-frame UV (280 nm), regardless of their star-formation activity, because evolved stellar populations produce a modest amount of UV light that traces the relatively smooth, symmetric, and centrally peaked stellar distribution seen in the near infrared. 
Ultraviolet morphologies among the BCGs with strong UV excesses exhibit distinctive knots, multiple elongated clumps, and extended filaments of emission that distinctly differ from the smooth profiles of the UV-quiet BCGs. 
These structures, which are similar to those seen in the few star-forming BCGs observed in the UV at low redshift, are
suggestive of bi-polar streams of clumpy star formation, but not of spiral arms or large, kpc-scale disks. 
Based on the number of streams and lack of culprit companion galaxies, these streams are unlikely to have arisen from multiple collisions with gas-rich galaxies.
These star-forming UV structures are morphologically similar to the cold-gas structures produced in simulations of precipitation-driven AGN feedback in which jets uplift low-entropy gas to greater altitudes, causing it to condense.
Unobscured star-formation rates estimated from CLASH UV images using the Kennicutt relation range up to 80 $M_\odot \, {\rm yr}^{-1}$ in the most extended and highly structured systems.
The circumgalactic gas-entropy threshold for star formation in CLASH BCGs at $z \sim 0.2$--0.5 is indistinguishable from that for clusters at $z < 0.2$. 
\end{abstract}


\section{Introduction}

The brightest cluster galaxy (BCG) is likely to be the dominant galaxy associated with a cluster of galaxies. 
As such, it is often but not always found near the bottom of the cluster's gravitational potential well \citep[e.g., ][]{2014ApJ...797...82L} and experiences a formation and interaction history unique to its special position in the hierarchy of galaxies \citep{2007MNRAS.375....2D}. 
Many BCGs exhibit colors and light profiles similar to those of quiescent elliptical galaxies, but a significant fraction have extremely extended halos, with Sersic indices \citep{1968Sersic} trending closer to $n \sim 10$ than to the canonical $n=4$ expected for elliptical galaxies \citep{1993MNRAS.265.1013C}. 

The unique features and relative rarity of BCGs in typical galaxy samples have arguably led to some neglect in studies of galaxy evolution. 
However, they provide an excellent opportunity to study the relationships between star formation, the circumgalactic medium (CGM), and feedback from a central active galactic nucleus (AGN).  
Baryons in the CGM are the probable gas supply for star formation in galaxies and possibly also for supermassive black-hole growth.
Around BCGs the CGM is hot enough to study in detail with X-ray observations, whereas the cooler CGM of less massive galaxies is probed only sparsely with UV and optical absorption-line studies along lines of sight to bright quasars \citep[e.g., ][]{1994ApJ...437L..75S,2013ApJ...777...59T,2014ApJ...792....8W}.

Multi-wavelength observational programs have shown that many BCGs are not quiescent systems and
exhibit activity such as radio emission ($>10^{23}$ W/Hz) \citep{1990AJ.....99...14B}, extended ($\sim10$ kpc)
emission-line nebulae \citep{1985ApJS...59..447H,1989ApJ...338...48H}, significant excess blue or UV light \citep{1998MNRAS.298..977C,2008ApJ...687..899R,2010ApJ...715..881D}, far-infrared emission from warm dust and PAHs \citep{2008ApJS..176...39Q,2011ApJ...732...40D}, and vibrationally excited molecular hydrogen \citep{2000ApJ...545..670D}.    This enhanced activity in the BCG is strongly correlated with the thermodynamic state of the hot CGM surrounding the BCG \cite[e.g., ][]{2008ApJ...683L.107C,2008ApJ...687..899R,2009ApJS..182...12C,2009ApJ...704.1586S}. 
In rich clusters of galaxies, all such active BCGs are embedded in cores or coronae of dense, low-entropy, high-pressure gas, commonly called ``cool cores" and formerly known as ``cooling flows"  \citep{2004cgpc.symp..143D}.  
The radiative cooling time of the hot gas surrounding active BCGs is typically $\lesssim 1$~Gyr. 

While 10-30\% of BCGs in optically-selected samples exhibit either radio-loud AGN or star-formation activity or both \citep[e.g.,][]{2007MNRAS.379..867V,2007MNRAS.379..100E}, the fraction rises to $>70\%$ in the local sample of X-ray selected clusters \citep{1999MNRAS.306..857C} and in a representative sample of X-ray clusters \citep{2010ApJ...715..881D}. 
Some of these BCGs have among the highest SFRs in the $z<0.6$ universe \citep[e.g.,][]{2006ApJ...648..164M,2013ApJ...765L..37M}. 
  The relationships observed between molecular gas content and star-formation rate in star-forming BCGs 
are remarkably similar to those observed in  star-forming galaxies and starbursts. 
Molecular gas ``depletion times" based on molecular hydrogen (H$_2$) masses derived from CO line luminosities  \citep{2001MNRAS.328..762E} and star-formation rates derived from FIR emission   are $\sim10^9$ years for both star-forming galaxies   and BCGs with SFRs of a few solar masses per year, 
and drop to $\sim 10^7 - 10^8$ years for both starburst galaxies  and BCGs with SFRs of tens or more solar masses per year \citep{2011ApJ...738L..24V,2008ApJ...681.1035O}.

One hypothesis for the link between a low-entropy CGM and activity in the BCG is that the lowest entropy gas in the CGM cools and provides cold gas to fuel both star formation and the AGN. 
A long-standing mystery is that unmoderated radiative cooling of the ICM would produce star formation rates at least an order of magnitude higher than observed \citep[e.g., ][]{1991A&ARv...2..191F}, hence the ``cooling flow'' problem. 
Spectroscopic evidence from the XMM-Newton grating spectrometer conclusively demonstrated that the ICM is not 
radiatively cooling from X-ray  emitting temperatures \citep{2003ApJ...590..207P}, leading most to assume that some kind of heating process involving the AGN nearly balances radiative cooling  
\citep{2007ARA&A..45..117M,2012NJPh...14e5023M}.
A second mystery is that, despite the strong correlation between low-entropy hot gas and cold, molecular, star-forming gas, it is not at all clear how hot, dust-free, X-ray emitting gas becomes dusty, cold molecular gas
\citep{2011ApJ...738L..24V}.

Observations of BCGs and the surrounding CGM suggest that the relationships between the AGN, its kinetic energy output, the BCG's star-formation activity, and the thermodynamic state of the intergalactic gas must be fairly cozy. Simulations of the broader population of galaxies indicate that star formation in the most massive galaxies must somehow be quenched by AGN feedback, because without it massive galaxies are not only much more luminous than galaxies we see today, but bluer as well, owing to high levels of continuous star formation \citep[e.g.,][]{Saro_2006.MNRAS.373..397S}. 
This problem is rectified in simulations that incorporate AGN feedback, indicating that AGNs are the likely culprits quenching continuous star formation in massive galaxies. 
The phenomena we observe in cool-core BCGs are likely to be signposts of AGN feedback in action. 
These AGN do not eliminate star formation, but are energetically capable of preventing the extreme star formation that would ensue if the ICM were allowed to cool unabated into molecular clouds \citep[e.g.,][]{Fabjan_2010.MNRAS.401.1670F,Dubois_2013.MNRAS.428.2885D,Ragone-Figeroa_2013.MNRAS.436.1750R,Martizzi_2014.MNRAS.443.1500M}.
 
 We and others have recently proposed a framework for precipitation-driven AGN feedback that account for these tight relationships 
\citep[e.g.,][]{2012MNRAS.419.3319M,2012MNRAS.420.3174S,2012ApJ...746...94G,2015ApJ...799L...1V,2015-Voit-Nature-arXiv1409.1598V}.
In this framework, radiative cooling of thermally unstable gas causes condensation and
precipitation of cool clouds out of the hot CGM, which then feed star formation and fuel growth 
of the central supermassive black hole.
The criterion for precipitation, indicated by numerical simulations of feedback and radiative cooling
in a spherical geometry is for the cooling time to be shorter than about ten times the free-fall time.
In the highest mass galaxies, precipitation causes a jump in jet power from the AGN, which subsequently 
limits cooling. 
The jets are also important for lifting low-entropy gas from the BCG's center out to larger radii where
it becomes unstable to condensation and creates filaments and knots of cooler gas.
If dust grains released by the BCG's old stellar population can survive until they are uplifted and 
incorporated into the condensing gas, then they can help to promote cooling and can also
serve as nucleation sites for converting a greater proportion of refractory elements into solid form.
Numerical simulations have begun to reproduce observable features of this physical model 
\citep[e.g. ][]{2014ApJ...789...54L,2014ApJ...789..153L}.

Here we present UV images from the Hubble Space Telescope Multi-Cycle Treasury program CLASH  \citep{2012ApJS..199...25P} that illustrate the most recent locations of star formation in active BCGs. 
Very few high-resolution UV images exist for brightest cluster galaxies, and the handful available at $z<0.3$ have revealed filamentary, clumpy structures \citep{1999ApJ...525..621K,2004ApJ...612..131O,2010ApJ...719.1619O,2010ApJ...719.1844H,2011MNRAS.414.2309O}. 
CLASH has substantially boosted the scant high-resolution UV coverage of BCGs by collecting high-quality images of 25 clusters of galaxies at $0.2 < z < 0.9$ through 16 filters spanning the ultraviolet to the near-infrared.  The rest-frame UV images of these galaxy clusters have no precedent in sensitivity or spatial resolution at similar redshifts.  We briefly describe the CLASH cluster sample in \S2. We describe the data analysis of the UV images for the 25 CLASH BCGs, estimate unobscured UV star-formation rates, and analyze UV morphology in \S3. We discuss these results and compare the UV morphologies to simulations of star formation in central cluster galaxies experiencing AGN-jet feedback in \S4.  We summarize our results in \S5.   

Throughout this paper we assume cosmological parameters of $\Omega_M=0.3$, $\Omega_\Lambda=0.7$, and $H_0=70 ~h_{70}$ km s$^{-1}$ Mpc$^{-1}$.

\section{CLASH}

The Cluster Lensing and Supernova with Hubble (CLASH) Multi-Cycle Treasure (MCT) program \citep{2012ApJS..199...25P} 
utilized 525 orbits (over $\sim500$ hours of observing time) from 2011--2013
(HST Proposal ID 12456). The sample includes 
25 clusters of galaxies with $z=0.2-0.9$ and global X-ray temperatures greater than 5--6 keV. Twenty of these
clusters were chosen to be examples of relaxed, massive systems, characterized by regular X-ray morphology on radial scales
greater than a few hundred kpc, with BCGs that are well-centered and aligned with the X-ray centroids. A significant number of these
clusters were included in the \citet{2008MNRAS.387.1179M} study of relaxed X-ray systems. The remaining 
five clusters were added to the sample 
because of their lensing properties and ability to magnify the high-redshift universe \citep{2012Natur.489..406Z,2013ApJ...762...32C}.
All of the CLASH clusters have Chandra observations, and 15 were also observed with XMM \citep{2014ApJ...794..136D}. 

One of the unprecedented features of the CLASH study was its use of 16 broadband filters, spanning UV to near-IR wavelengths. 
The motivation for this broad coverage was to significantly improve the photometric redshift estimates of lensed galaxies, but a side benefit is that the cores of these clusters received UV exposure otherwise rather difficult to obtain.
Only a few of the CLASH BCGs had published UV observations prior to CLASH. A383, m1115,
r2129, and ms2137 have GALEX-based SFR upper limits of $<2-3~\rm{M}_\odot~\rm{yr}^{-1}$. Only
one, m1720, had a published estimate of its UV star-formation rate 
($2.7 \pm 0.7~\rm{M}_\odot~\rm{yr}^{-1}$; Hoffer et al. 2012). 

 These UV images provide our first high resolution view of a sample of BCGs in massive clusters at moderate redshifts, but many of them have been the subject of studies at other wavelengths, only a few of which we have space to mention here.
Several of the BCGs are known to be mid-IR sources of dust emission, some of which can be produced by radiative heating of dust by light from evolved stars, and some show clear evidence for recent star formation.
Four CLASH BCGs have mid-IR detections with Spitzer (a209, A383, ms2137, r2248 and r1532), and there are mid-IR upper limits for several more (c1226,  A2261, and r2129). The most extreme star-forming system known in CLASH prior to this study was r1532, with a Spitzer-based SFR of $110 ~\rm{M}_\odot~\rm{yr}^{-1}$ \citep{2012ApJS..199...23H}.  The most extreme AGN-influenced BCG is arguably m1931 \citep{2011MNRAS.411.1641E}.  
Uniformly derived profiles of temperature, ICM electron density, gas entropy, gas mass, and total mass based on analysis of the pressure profiles are provided in \citet{2014ApJ...794..136D} for all 25 of the CLASH clusters.  Gravitational
lensing profiles for a significant fraction of the CLASH sample are published  \citep{2014ApJ...795..163U,2014arXiv1404.1376M,2014arXiv1411.1414Z}.
 All of these sources have either upper limits or detections from radio surveys, which we will discuss in \S~\ref{agn}.

\section{CLASH Observations}

The CLASH clusters, mean redshifts, observation dates, and filter names are given in Table~\ref{table:obs}. Filter names are provided using the convention that an HST filter named FnnnW is listed as ``nnn" in the table. For HST,
the filters shortward of 1 micron are named for the approximate central wavelength of the filter in nanometers; filters
longward of 1 micron (which we call infrared or IR for short) are named for this wavelength in units of 10 nanometers (so 1.6 micron would correspond to 160 on this scale).

\subsection{HST Data Reduction and Photometry}  
  
The HST ACS and WFC3 imaging data for these clusters were processed through initial detector-level calibration to 
remove instrumental signatures from each exposure. Exposures in all filters were subsequently aligned onto a 
uniform astrometric grid to an accuracy better than $\sim\,$2-3~milliarcseconds, passed through cosmic-ray and 
bad-pixel rejection, and combined into final mosaics for each filter at pixel scales of 30~milliarcseconds and 
65~milliarcseconds.  All of this was done using a modified version of the MosaicDrizzle pipeline 
\citep{2011ApJS..197...36K, 2002hstd.book.....K}. 
Further details are presented in the CLASH overview paper \citep{2012ApJS..199...25P}.
We used mosaics binned to 65~milliarcseconds per pixel for this study. For the WFC3 UV images,
later versions of our mosaics were significantly improved over the first version of publicly available 
mosaics because of better flat-fielding and sky treatment.  For the sake of reproducibility, 
we provide the mosaic production date in Table~\ref{table:obs}.

\begin{deluxetable*}{llrcc|cc|cc|cll}
\tablecaption{HST Observations and Processing Dates \label{table:obs}}
\tablehead{ \colhead{Full Name} &
	\colhead{Cluster} & \colhead{z} & \colhead{ObsDate (UV)} & \colhead{ProcDate} & \colhead{UV\tablenotemark{1}} & \colhead{$t_{exp}$}
	& \colhead{NIR\tablenotemark{1}} & \colhead{$t_{exp}$}& \colhead{ObsDate (NIR)} & \colhead{ $A_{UV}$} & \colhead{ $A_{NIR}$ } \\
	\colhead{} &
	\colhead{} & \colhead{} & \colhead{} & \colhead{} & \colhead{} & \colhead{(s)}
	& \colhead{} & \colhead{(s)}& \colhead{} & \colhead{mag} & \colhead{mag}} 
\startdata
Abell 1423 & A1423   & 0.213 & 12/28/12  & 2/27/13    & 336      & 4890  & 125      & 2514  & 12/31/12  & 0.085 & 0.014 \\
Abell 209  & A209    & 0.206 & 9/9/12    & 1/19/13    & 336      & 4752  & 125      & 2514  & 8/12/12   & 0.086 & 0.014 \\
Abell 2261 & A2261   & 0.224 & 4/3/11    & 1/23/13    & 336      & 4817  & 125      & 2514  & 3/9/11    & 0.193 & 0.031 \\
Abell 383  & A383    & 0.187 & 12/28/10  & 7/2/11     & 336      & 4868  & 125      & 3320  & 1/5/11    & 0.14  & 0.023 \\
Abell 611  & A611    & 0.288 & 3/1/12    & 6/18/12    & 336      & 4782  & 125      & 2514  & 1/28/12   & 0.255 & 0.042 \\
WARP 1226.9+3332 & c1226 & 0.890 & 5/24/13   & 6/26/13    & 475      & 4396  & 160      & 5129  & 5/9/13    & 0.063 & 0.014 \\
MACSJ0329.6-0211 & m0329   & 0.450 & 9/26/11   & 11/5/11    & 390      & 4902  & 160      & 5029  & 8/18/11   & 0.235 & 0.031 \\
MACSJ0416.1-2403 & m0416   & 0.395 & 8/18/12   & 1/19/13    & 390      & 4814  & 160      & 5029  & 7/24/12   & 0.16  & 0.021 \\
MACSJ0429.6-0253 & m0429   & 0.399 & 12/11/12  & 2/27/13    & 390      & 4562  & 140      & 2312  & 12/5/12   & 0.235 & 0.037 \\
MACSJ0647.8+7015 & m0647   & 0.584 & 11/16/11  & 12/2/11    & 435      & 4248  & 160      & 5229  & 10/22/11  & 0.402 & 0.057 \\
MACSJ0717+3745   & m0717   & 0.548 & 10/29/11  & 12/30/11   & 435      & 4052  & 160      & 5029  & 10/10/11  & 0.277 & 0.039 \\
MACSJ0744.9+3927 & m0744   & 0.686 & 11/17/11  & 12/30/11   & 475      & 4022  & 160      & 5029  & 10/29/11  & 0.189 & 0.03  \\
MACSJ1115.8+0129 & m1115   & 0.355 & 1/31/12   & 3/5/12     & 390      & 4554  & 140      & 2012  & 1/7/12    & 0.151 & 0.024 \\
MACSJ1149.6+2223 & m1149   & 0.544 & 2/13/11   & 3/14/11    & 435      & 3976  & 160      & 5029  & 1/16/11   & 0.083 & 0.012 \\
MACSJ1206.2-0847 & m1206   & 0.439 & 5/25/11   & 8/15/11    & 390      & 4959  & 160      & 5029  & 1/16/11   & 0.246 & 0.032 \\
MACSJ1311.0-0311 & m1311   & 0.494 & 6/12/13   & 7/9/13     & 435      & 4172  & 160      & 2011  & 4/22/13   & 0.112 & 0.016 \\
MACSJ1423.8+2404 & m1423   & 0.545 & 2/5/13    & 3/13/13    & 435      & 4196  & 160      & 5029  & 1/19/13   & 0.099 & 0.014 \\
MACSJ1720.2+3536 & m1720   & 0.387 & 5/22/12   & 6/28/12    & 390      & 4810  & 140      & 2312  & 5/9/12    & 0.147 & 0.023 \\
MACSJ1931.8-2635 & m1931   & 0.352 & 5/28/12   & 6/28/12    & 390      & 4620  & 140      & 2312  & 4/14/12   & 0.431 & 0.068 \\
MACSJ2129-0741   & m2129   & 0.570 & 6/24/11   & 1/22/12    & 435      & 3728  & 160      & 5029  & 5/15/11   & 0.273 & 0.039 \\
MS 2137.3-2353   & ms2137  & 0.313 & 9/29/11   & 1/21/13    & 390      & 4827  & 140      & 2312  & 9/9/11    & 0.198 & 0.031 \\
RXJ1347.5-1145   & r1347   & 0.451 & 6/22/11   & 8/12/11    & 390      & 4820  & 160      & 7741  & 5/21/10   & 0.241 & 0.032 \\
MACSJ1532.8+3021 & r1532   & 0.363\tablenotemark{2} & 3/29/12   & 4/16/12    & 390      & 4840  & 140      & 2312  & 3/16/12   & 0.115 & 0.018 \\
RXJ2129.6+0005   & r2129   & 0.234 & 5/31/12   & 10/16/12   & 336      & 4574  & 125      & 3420  & 4/3/12    & 0.18  & 0.029 \\
ABELL S1063      & r2248   & 0.348 & 9/24/12   & 1/19/13    & 390      & 4740  & 140      & 2312  & 9/12/12   & 0.048 & 0.007
\enddata
\tablenotetext{1}{Filter names are listed here in a compressed format, such that nnn represents filter FnnnW for the ACS or WF3.}
\tablenotetext{2}{Redshift of this cluster and its BCG is based on the SDSS emission-line spectrum of the BCG.}
\end{deluxetable*}

Rest-frame UV was defined to be 280 nm in the rest-frame of the BCG. At this wavelength, the UV from recently formed
hot stars, if present, readily exceeds the light emitted by older stars. At shorter wavelengths, the relative 
contribution of ``UV-upturn" stars increases below 250 nm \citep[e.g.,][]{2004Ap&SS.291..215B} 
with considerable variation from one galaxy to the next, 
even in a sample of quiesent ellipticals  \citep[e.g.,][]{2005ApJ...619L.107R,2009ApJ...694.1539A}.
To allow an estimate of the level of the evolved stellar population, we chose the image obtained  
at a rest wavelength of $\sim 1$ micron, chosen because 1 micron is approximately the peak of the stellar-light spectrum for passive galaxies and evolved stars. 
For the highest redshift CLASH clusters, c1226 and m0744, that wavelength was not
in the CLASH coverage, so the near-IR (NIR) photometry is based on the reddest bandpass available (F160W). 
Resolved dust maps, emission-line maps, and optical long-slit 
spectroscopy will be presented in a future work (Fogarty et al., in prep.)
For accurate color photometry, the UV images were degraded with a  Gaussian
kernel to match the PSF in the NIR prior to photometry measurements. 

We measured the total UV emission from each BCG using the IRAF 2.15.1a package {\em apphot} 
within a circular aperture centered on the BCG\footnote{\url{http://iraf.noao.edu}}, 
with coordinates and radii listed in Table~\ref{table:phot}. 
For BCGs with structures visible in the UV, the aperture size was chosen to provide an estimate of the total amount of
UV light from the highest surface brightness structures in the BCG. 
If the field was uncrowded and the UV source centrally concentrated and extended, 
a default aperture of $14$ kpc (equal to $10/h$ kpc) was chosen to estimate the color of quiescent BCGs.
However, this aperture size was reduced in crowded fields to avoid including light from neighboring galaxies, or in the 
case of m1149, a prominent and extended lensed star-forming background galaxy.
 UV and 1.0-micron photometry are based on identical apertures for the same galaxy. 

\begin{deluxetable*}{llllllllllll}
\tablecaption{UV and NIR Photometry \label{table:phot}}
\tablehead{
\colhead{Cluster}      & \colhead{Radius}      & \colhead{Radius}      & \colhead{UV (AB)} & \colhead{$\pm$} & \colhead{UV-IR}   & \colhead{$\pm$} & \colhead{Log L$_{UV}$}    & \colhead{SFR}   & \colhead{SFR} & \colhead{SFR}  & \colhead{$K_0$\tablenotemark{c}}    \\
 \colhead{}              & \colhead{arcsec}  & \colhead{kpc}          & \colhead{mag} & \colhead{mag} & \colhead{mag} & \colhead{mag }     & \colhead{erg s$^{-1}$ Hz$^{-1}$  }       & \colhead{M$_\odot$ yr$^{-1}$} & \colhead{corr} & \colhead{err} & \colhead{keV cm$^2$}     }
\startdata
m1931   & 5.75          & 28.5        & 18.52  & 0.03          & 2.04  & 0.04     & 29.77 & 83.08    & 78.12     & 2.34     & $14\pm 4$  \\
r1532   & 7.02          & 35.5        & 19.18  & 0.03          & 2.83  & 0.04     & 29.54 & 48.62    & 42.61     & 2.64     & $17\pm 2$  \\
m0329   & 3.25          & 18.7        & 20.22  & 0.02          & 3.3   & 0.03     & 29.35 & 30.99    & 25.08     & 2.39     & $11\pm 3$  \\
m0429   & 3.90          & 20.9        & 20.38  & 0.04          & 3.75  & 0.05     & 29.16 & 20.11    & 14.31     & 2.07     & $17\pm 4$  \\
m1423   & 1.20          & 7.7         & 21.39  & 0.01          & 3.14  & 0.02     & 29.08 & 16.70    & 13.95     & 1.15     & $10\pm 5$  \\
r1347   & 2.50          & 14.4        & 20.91  & 0.02          & 3.81  & 0.03     & 29.07 & 16.50    & 11.47     & 1.75     & $12\pm 20$  \\
m1115   & 1.53          & 7.6         & 21.33  & 0.01          & 3.38  & 0.02     & 28.66 & 6.37     & 5.06      & 0.52     & $15\pm 3$  \\
ms2137  & 1.56          & 7.2         & 21.16  & 0.02          & 4.07  & 0.03     & 28.60 & 5.56     & 3.41      & 0.66     & $15\pm 2$  \\
m1720   & 2.70          & 14.2        & 21.6   & 0.05          & 4.54  & 0.05     & 28.64 & 6.08     & 2.45      & 0.74     & $24\pm 3$ \\
A383    & 2.30          & 7.2         & 20.46  & 0.03          & 4.36  & 0.04     & 28.37 & 3.29     & 1.63      & 0.41     & $13\pm 2$ \\
r2129   & 3.80          & 14.2        & 21.16  & 0.09          & 4.98  & 0.09     & 28.31 & 2.86     & \nodata      & $<0.4$     & $21\pm 4$  \\ \hline
m0744   & 2.00          & 14.2        & 22.73  & 0.13          & 4.6   & 0.13     & 28.78 & 8.50     & \nodata      & $<3.1$     & $42\pm 11$  \\ 
m1311   & 2.41          & 14.6        & 22.29  & 0.14          & 4.69  & 0.14     & 28.61 & 5.75     & \nodata      & $<1.9 $    & $47\pm 4$  \\
A2261   & 3.93          & 14.2        & 20.9   & 0.06          & 5.47  & 0.07     & 28.37 & 3.29     & \nodata     & $<2.8$     & $61\pm 8$  \\
m1206   & 2.50          & 14.2        & 21.81  & 0.04          & 4.47  & 0.05     & 28.68 & 6.75     & \nodata     & $<3$\tablenotemark{a}     & $69 \pm 10$  \\
A1423   & 4.10          & 14.2        & 21.22  & 0.13          & 4.96  & 0.13     & 28.19 & 2.19     & \nodata      & $<0.4$     & $68\pm 13$  \\
a209    & 2.50          & 8.5         & 21.79  & 0.11          & 5.48  & 0.11     & 27.93 & 1.20     & \nodata     & $<1.1$     & $106\pm 27$ \\
A611    & 1.69          & 7.3         & 22.93  & 0.13          & 5.69  & 0.14     & 27.81 & 0.90     & \nodata     & $<1.7$     & $125\pm 18$ \\
c1226 & 0.78            & 6.1         & 24.68  & 0.17          & 5.37  & 0.17     & 28.28 & 2.67     & \nodata     & $<1.5$     & $166\pm 45$ \\
r2248   & 1.24          & 6.1         & 22.39  & 0.04          & 4.91  & 0.04     & 28.21 & 2.29     & \nodata      & $<0.5$     & $170 \pm 20$ \\ \hline
m0416   & 2.60          & 13.9        & 22.26  & 0.09          & 5.24  & 0.10     & 28.39 & 3.48     & \nodata     & $<0.8$     & $400\pm 100$ \\
m0647   & 1.11          & 7.3         & 23.8   & 0.29          & 5.19  & 0.29     & 28.19 & 2.14     & \nodata     & $<0.3$     & $225\pm 50$ \\ 
m0717\tablenotemark{b}  & 1.95 & 12.5 & 22.63  & 0.15          & 4.85  & 0.15     & 28.59 & 5.40     & \nodata      & $<1.4$     & $220\pm 96$ \\
m1149   & 1.11          & 7.0         & 23.63  & 0.15          & 4.73  & 0.15     & 28.18 & 2.11     & \nodata      & $<0.7$\tablenotemark{a}     & $280\pm 40$ \\
m2129   & 1.10          & 7.2         & 24.03  & 0.22          & 5.07  & 0.22     & 28.07 & 1.64     & \nodata      & $<0.1$     & $200 \pm 100$ \\

\enddata
\tablecomments{The CLASH clusters with $K_0 < 30$ keV cm$^{2}$ are in the top section, ordered by estimated UV star formation rate, from
large to small. The next section are clusters with $K_0 > 30$ keV cm$^{2}$, ordered by estimated core entropy $K_0$. 
The bottom section lists the five high magnification clusters from CLASH, in order of right ascension. All of the information in this table
is from this work except for the $K_0$ estimates, which are primarily from ACCEPT \citep{2009ApJS..182...12C}, except for m0416 and m2129 \citep{2014ApJ...794..136D}.}
\tablenotetext{a}{The clusters m1206 and m1149 both have known lensed background systems that contaminate the UV images of their BCGs. The upper limit on 
excess UV in m1206  associated with star formation or other BCG activity is set equivalent to the level of detected excess, while the aperture used for m1149 avoids the lensed galaxy. }
\tablenotetext{b}{The cluster m0717 does not have a prominent BCG or even a double BCG. We have identified a bright candidate BCG with a spectroscopic redshift
consistent with being a cluster member, as discussed in the text. Coordinates for all BCGs, along with GALEX photometry for comparison, are provided in Table~\ref{table:GALEX_NVSS}}
\tablenotetext{c}{ The estimated central entropies  $K_0 = T_X/n_e^{2/3}$ in units of keV cm$^{2}$ are from ACCEPT \citep[][]{2009ApJS..182...12C}. For the non-ACCEPT cluster m0416, the $K_0$ is based on electron density and temperature profiles from the X-ray analysis described in Donahue et al. (2014). M0416 is an interacting cluster, so the high gas entropy in the center is not surprising.}
\end{deluxetable*}

To investigate the effects of large-scale scattered light, we obtained UV background measurements 
in annuli of multiple sizes surrounding the aperture. 
Robust UV photometry measurements are optimized with an inner radius for the background 
annulus of about $15\arcsec$, although
the result is fairly insensitive to choices smaller and larger than this.
We further investigated background systematics by randomly moving the aperture/annulus combination 
for each cluster, avoiding  sources of obvious emission or scattered light. We then performed aperture photometry on these blank pieces of sky in the same manner as on the BCG. From these measurements, we found the standard  deviation in flux, using a 3-$\sigma$ cut to exclude possibly real but low-flux UV sources. We use this  standard deviation as the systematic uncertainty in our photometric measurements resulting from  scattered light in the field and other flat fielding issues. 

The NIR sky background was estimated in an aperture well outside the brightest part of the BCG to
give our best estimate of the ambient sky. The background sky levels were 
based on a median estimate with 3-$\sigma$ clipping.
After aperture fluxes were converted to AB magnitudes using the ACS and WFC3 
zeropoints,\footnote{\url{http://www.stsci.edu/hst/acs/analysis/zeropoints/zpt.py}, \url{http://www.stsci.edu/hst/wfc3/phot_zp_lbn}}
all photometry was corrected for Galactic extinction at the observed bandpass (Table~\ref{table:obs}), ranging 
from $0.08-0.45$ mag in the UV and  $<0.01 - 0.07$ mag in the NIR, using HST bandpass-specific extinction corrections  assuming $R_V=3.1$ and based on \citep{1999PASP..111...63F}based on \citet{2011ApJ...737..103S}, which is a recalibration of \citet{1998ApJ...500..525S}, and.

We estimated K-corrections based on the assumption that the NIR spectrum is dominated by a 10 Gyr old stellar population, which was modeled with HST bandpasses convolved with Starburst99 models \citep{1999ApJS..123....3L}. 
The estimated K-correction is less than 0.1 magnitudes for all the clusters and 
does not affect our results in an interesting way. 
We did not attempt to homogenize the UV photometry for small offsets from the nominal rest-frame wavelength of 280 nm, since
the UV light from a population of hot stars is nearly flat in $F_\nu$. We verified this assumption by comparing the 280
nm rest-frame photometry with bluer data, when available.

Statistical errors for both UV and NIR were computed by {\em apphot} based on counting statistics, obtained
by setting the appropriate electron-to-count-rate ratio equal to the effective gain, 
and the standard deviation in the background annuli. (The mosaics in the CLASH 
pipeline have the units of electrons sec$^{-1}$, so to compute the appropriate counting statistics for
a given mosaic, the gain parameter in the {\em apphot} task was set to the exposure time.)
Because the galaxies are so bright in the NIR in the
aperture of interest, the exact NIR background and its uncertainty are negligible for this work. 

Systematic and statistical  flux errors were added in quadrature to estimate the $1\sigma$ uncertainties given in Table 2. Color uncertainties include an estimated calibration uncertainty of 2\%, to account for uncertainties in absolute calibration, relative calibration between the two filters, and the unknown underlying spectral shape compared to that assumed to estimate the AB magnitude at a given wavelength.

\subsection{UV Star-Formation Rates}
  
In order to estimate the UV light from recent star formation, we first need to account for the UV
light contributed by the evolved stellar population. For this purpose, we use the quiescent
BCGs as templates. The fact that the BCGs are at different redshifts
introduces some scatter into our estimate, but not enough to affect these estimates significantly.
This method has the advantage of obtaining the UV contribution from evolved stars from an otherwise similar population of galaxies: the BCG population itself. 

The UV-NIR colors of quiescent  BCGs are fairly well behaved within the CLASH sample. 
We find an average 280-1000 micron color of $5.13 \pm 0.35$ (omitting m1206 because of contamination by
a lensed galaxy).
This is consistent with the mean and dispersion of quiescent BCGs in other studies, and
a baseline (established by the 3 reddest BCGs) of $5.5 \pm 0.5$. 
In \citet{2012ApJS..199...23H}
we found that the typical UV--K color of relatively nearby, quiescent BCGs was $6.6 \pm 0.3$, and typical $J-K$ colors are $\sim 1$. The BCGs with the faintest inferred UV luminosities also have
the reddest UV--1.0 micron colors of $\sim 5-6$, consistent with UV--NIR estimates from \citet{2012ApJS..199...23H,2010ApJ...719.1844H,2010ApJ...715..881D}. 
We use the average color
to correct for the UV produced by the evolved stars in the BCG.  Using a redder color would increase
our SFR estimates most significantly for the BCGs with the lowest rates. 

We convert excess UV luminosity to an approximate 
star-formation rate using the UV conversion from \citet{1998ARA&A..36..189K}. 
We note that the Kennicutt SFR conversion is based on the \citet{1955ApJ...121..161S} initial mass function (IMF) and the assumption of 
a nearly constant star formation rate over the last 100 million years or so. There may be differences in the low-mass end of the 
IMF between BCGs and the star-forming galaxies studied by Kennicutt \citep[see e.g.,][]{2010Natur.468..940V}, but for
our purposes her the relative SF rates are more relevant than the absolute rates. 
For the BCGs which may be relatively young starbursts, these estimated UV rates underestimate the unabsorbed SFR. We have
not attempted to estimate the star formation that is obscured by dust clouds, but typically, mid-IR based SFR are similar to or larger than the unobscured SFRs (e.g., Hoffer et al. 2012). 
 
We find that of the 25 CLASH clusters, seven exhibit unambiguous UV excesses, corresponding to unabsorbed star-formation rates of 5--80 solar masses per year (m1931, r1532, m0329, m0429, m1423, r1347, 
m1115). If we adopt a detection threshold of 3 sigma, three other clusters
exhibit UV excesses possibly associated with star formation (A383, m1720, and ms2137).  Of these three, only ms2137 fails to show morphological evidence for a young stellar population clearly distinct from the evolved stellar population and could contain a low-luminosity AGN. 

Four of the CLASH BCGs have Herschel mid-IR detections from \citet{2012ApJ...747...29R},
from which they derive SFRs of $46 \pm 3$, $4 \pm 0.2$, $1.6 \pm 0.1$, and $1.7 \pm 0.1$ M$_\odot$ yr$^{-1}$
for m1423, A383, r1720, and r2129, respectively. These quantities are quite similar to our UV
estimates and limits. We did not detect excess UV emission from r2129, but distinguishing 
between excess UV from young stars and the UV emission from an evolved population 
becomes difficult with photometry alone at $\sim 1 \, M_\odot$ yr$^{-1}$. 
Ten other CLASH BCGs in \citet{2012ApJ...747...29R} were not detected in the mid-IR.
Among these IR non-detections, we find a UV excess for only ms2137.

In the \citet{2012ApJS..199...23H} compilation of SFRs, a209 has an estimated 
SFR of $0.9 \pm 0.1$ M$_\odot$ yr$^{-1}$, based on 24 micron photometry, while our measurements give an 
upper limit of $<1.1$ M$_\odot$ yr$^{-1}$. The Hoffer SFR for A383 of $1.6 \pm 0.2$ is similar to our UV rate 
of $1.6 \pm 0.4$ M$_\odot$ yr$^{-1}$. The UV-quiescent BCG in r2248 (also known as AS1036) has a small 
IR excess corresponding to $0.8 \pm 0.13$ M$_\odot$ yr$^{-1}$. As mentioned above,  r1532 is a known 
starbursting BCG with a mid-IR SFR of $110\pm 22$ M$_\odot$ yr$^{-1}$ and
a UV rate of $\sim 43$  M$_\odot$ yr$^{-1}$. In general, mid-IR observations around 100 microns, 
near the peak of the dust emission, are more sensitive to low levels of star formation or other activity because of the uncertain UV contribution from evolved stars. 
However, dust heated by light from the old stellar population can also produce FIR emission, so low-level SFRs for BCGs based on single IR photometry points can also be inaccurate \citep{2011ApJ...732...40D}. 
Ideally, one would like more accurate measures of star formation based on multiwavelength observations.
However, in most of the star-forming CLASH BCGs, we are not hampered by uncertainties in UV flux from 
the evolved population because those galaxies exhibit both UV excesses of clear statistical significance and
morphological indications that the bulk of the UV emission is not coming from the evolved stellar population.
 									
\subsection{UV Morphologies}

\begin{figure*}
\includegraphics[width=6.0in,trim = -1.5in 0.0in 0.0in 0.0in]{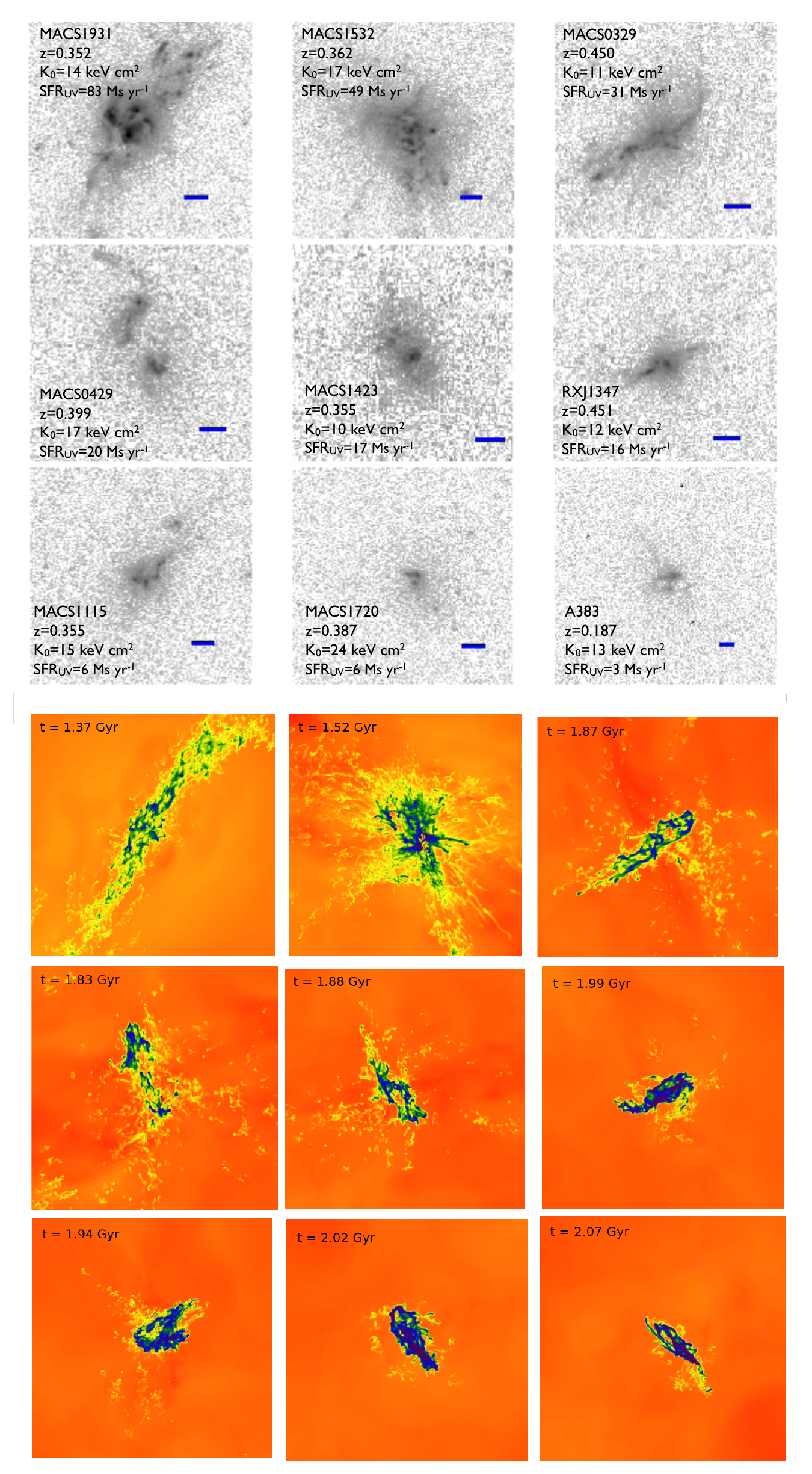}
  \caption{\footnotesize 
The top 9 images show the central 50~kpc$\times$50~kpc of the 280 nm broad-band HST images for star-forming CLASH BCGs. A blue bar on each frame indicates a $1\arcsec$ angular scale. 
Images are oriented with North up and East to the left. 
The bottom 9 images show gas-temperature maps with similar physical scales from a simulation of a single BCG in which precipitation of cold clouds out of the CGM triggers jets from the central AGN \citet{2014ApJ...789...54L,2014ApJ...789..153L}. Cold gas in the simulations (show in blue) displays structures similar in extent and axis ratio to the UV structures seen in the star-forming CLASH clusters. 
Time stamps in the simulation frames show the time in Gyrs since the beginning of the simulation. (The first major AGN outburst occurs at $\sim 0.3$ Gyrs.)
\label{UVimage1}}
\end{figure*}

\begin{figure*}
 \begin{minipage}[b]{1.0\linewidth}
\begin{center}
    \includegraphics[width=4.5in, trim = 4.0in 5.9in 4.3in 0.3in, clip=true]{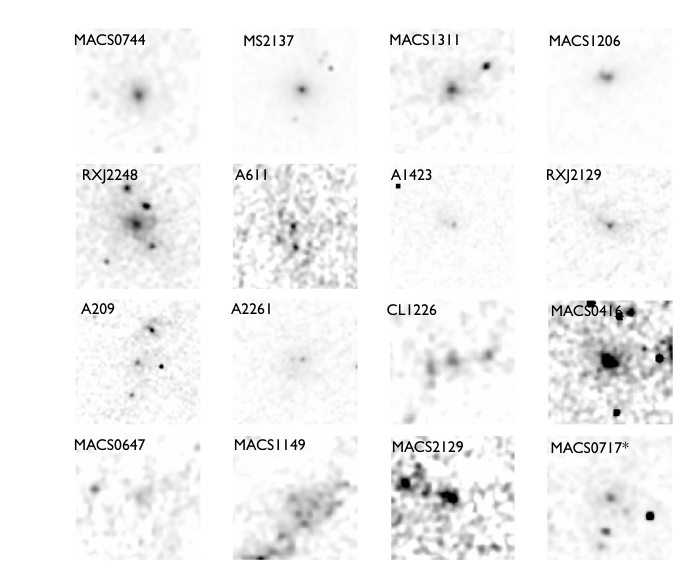}
\end{center}
  \end{minipage}
\caption{\footnotesize
Ultraviolet (280 nm rest-frame) broad-band HST images of the central 50~kpc~$\times$~50~kpc of 16 CLASH clusters with moderate to high entropy cores, smoothed with a gaussian having FWHM of 0.3".
Most of the BCGs in these clusters do not show much 2-d structure beyond a faint, central enhancement. 
Exceptions are the m1206 BCG, with a tadpole-shaped central feature that is likely to be the lensed light of a background galaxy, and the m1149 BCG which has a prominent, lensed spiral galaxy just SE of its core. 
All of the BCGs are centered in each image. 
In some cases the UV source is very compact, likely to be a nuclear star cluster or a weak AGN. 
Images are oriented with North up and East to the left. 
\label{UVimage2}}
\end{figure*}

We show raw (un-smoothed) UV images of the star-forming CLASH BCGs in Figure~\ref{UVimage1}, and smoothed UV images of the rest of the CLASH BCGs in Figure~\ref{UVimage2}.
Two of the clusters with quiescent BCGs, m1206 and m1149, have well-known lensed background 
features that contaminate UV estimates from the core of the cluster. m0744 and m1311 
are somewhat bluer than the average quiescent BCG, but they both show smooth isophotes and little morphological evidence for a distinct star-forming population.

In order to characterize the UV and IR morphologies of the BCGs in the 20 X-ray selected CLASH clusters 
we compute low-order moments of their surface brightness distributions. 
For each BCG we compute these moments inside a circular region that excludes other galaxies and estimate a scale and position angle from the resulting moment matrix. 
This exercise can be done analytically, so we provide the full procedure below for completeness.

The surface brightness centroid ($C_x$, $C_y$) is determined by the normalized sums $C_x = \sum i(x,y) x $
and $C_y = \sum i(x,y) y $, where $I(x,y)$ is the surface brightness of pixel $(x,y)$ and $i(x,y) \equiv I(x,y) /  \sum I(x,y)$. The sums are over all $(x,y)$ inside a circular aperture with a radius chosen to avoid excess off-axis contamination from other galaxies. Small variations in this aperture size do not change the result.  The matrix elements of the next moment of the surface brightness distribution are $M_{xx} = \sum i(x,y)*(x-C_x)^2  $, $M_{yy} = \sum i(x,y)*(y-C_y)^2  $ and $M_{xy}=M_{yx}=\sum i(x,y)*(x-C_x)(y-C_y)  $. These terms define a symmetric $2 \times 2$ matrix 
\begin{equation}
A = \begin{pmatrix} a & c \\ c & b \end{pmatrix}
\end{equation}
with $M_{xx}=a$, $M_{yy}=b$, and $M_{xy}=M_{yx}=c$. 
This matrix can be diagonalized in a standard way with a rotation corresponding to 
\begin{equation}
\tan \eta = \left[ \left(\frac{a-b}{c}\right)^2 + 1 \right]^{1/2} - \left( \frac{a-b}{2c} \right) \; \; ,
\end{equation}
giving the eigenvalues 
\begin{equation}
\lambda_{\pm} = \frac{a+b}{2} \pm \left[ \left( \frac{a-b}{2c}\right)^2 + 1 \right]^{1/2}
\end{equation}
as diagonal elements.
These elements represent one way to quantitatively estimate the extent of a complex region. We define the terms 
 $x_s = \sqrt{\lambda_+}$ and $y_s = \sqrt{\lambda_-}$. The position angle $\eta$ (PA in Table~\ref{table:morph}) is defined
 to be that of the semi-major axis, in degrees east of north (where north is aligned along the $y$ axis as a convention in our mosaics), and axis ratio is defined to be the minimum of $y_s/x_s$ or $x_s/y_s$. 
 
We computed these quantities for both the UV and NIR images and report the results in Table~\ref{table:morph}. 
The clusters that host BCGs with clear evidence for star formation
are listed first in this table, approximately in order of star-forming region color, from blue to red.
All of the surface brightness moments for the star-forming BCGs were computed within a region with a radius of about $5\arcsec$, 
while the moments computed for quiescent BCGs were restricted to regions of $1.5-3.5\arcsec$   
to avoid including an unnecessarily large number of sky- or readnoise-dominated pixels and to avoid confusion with unrelated UV sources.  However, the UV and NIR regions used to compute morphological quantities
for each cluster are exactly the same, so that comparison between results from the UV and the NIR from
cluster to cluster is relevant. The uncertainties for all quantities in this table were computed 
by calculating the standard deviation of morphological parameters using 100 bootstrapped images with gaussian noise proportional 
to the inverse of the square root of the mosaic weight matrix. The statistical uncertainties for properties 
derived from the IR image were negligible compared to those derived for the UV images.

\begin{deluxetable*}{lcccrrrrccc}
\tablecaption{UV and NIR Morphological Parameters \label{table:morph}}
\tablehead{
\colhead{Name}        & \colhead{Aperture} & \colhead{Centroid Diff\tablenotemark{b}} & \colhead{$\pm$} & \colhead{PA (UV)\tablenotemark{a}} & \colhead{PA (IR)\tablenotemark{a}}  & \colhead{PA Diff}\tablenotemark{b}  & \colhead{$\pm$} & \colhead{Axis Ratio} & \colhead{$\pm$}  & \colhead{Axis Ratio\tablenotemark{c}}    \\
\colhead{  }          & \colhead{(arcsec)}         & \colhead{(arcsec)}          & \colhead{(arcsec)}       & \colhead{(degrees)}    & \colhead{(degrees)} & \colhead{(degrees)} & \colhead{(degrees)}     & \colhead{(UV)}     & \colhead{ }    & \colhead{(IR)}     } 
\startdata 
m1931    & 4.88     & 0.83              & 0.01                  & -35     & -31.38  & 3.2           & 0.4         & 0.61          & 0.01              & 0.83          \\
r1532     & 4.88     & 0.06              & 0.01                  & 41      & 52.91   & 12.2          & 0.9         & 0.79          & 0.01              & 0.79          \\
m0329    & 4.88     & 0.23              & 0.02                  & -49     & -5.36   & 43.3          & 1.4         & 0.76          & 0.01              & 0.94          \\
m0429    & 4.88     & 1.12              & 0.02                  & 18      & -6.50   & 24.7          & 1.8         & 0.75          & 0.01              & 0.83          \\
m1423    & 4.88     & 0.09              & 0.04                  & 21      & 29.07   & 7.6           & 3.5         & 0.65          & 0.05              & 0.77          \\
r1347     & 4.88     & 0.24              & 0.03                  & -49     & -6.88   & 42.2          & 10.3        & 0.93          & 0.02              & 0.88          \\
m1115    & 4.88     & 0.33              & 0.04                  & -43     & -40.29  & 2.6           & 2.4         & 0.78          & 0.02              & 0.86          \\
ms2137      & 2.28     & 0.05              & 0.01                  & 71      & 61.08   & 10.4          & 6.7         & 0.96          & 0.01              & 0.97          \\
m1720    & 4.88     & 0.21              & 0.04                  & -37     & -15.49  & 21.8          & 34.0        & 0.97          & 0.06              & 0.87          \\
A383  & 4.88     & 0.12              & 0.05                  & 27      & 13.50   & 13.2          & 7.2         & 0.88          & 0.02              & 0.96          \\
r2129     & 2.28     & 0.19              & 0.03                  & 69      & 68.24   & 1.1           & 3.5         & 0.74          & 0.02              & 0.88          \\ \hline
m0744    & 2.28     & 0.19              & 0.03                  & 60      & 27.52   & 32.3          & 18.5        & 0.91          & 0.05              & 0.93          \\
m1311    & 2.28     & 0.20              & 0.04                  & -66     & -45.33  & 20.8          & 7.1         & 0.81          & 0.05              & 0.94          \\
A2261 & 2.28     & 0.18              & 0.03                  & -44     & -27.25  & 16.6          & 12.1        & 0.94          & 0.02              & 0.98          \\
m1206    & 4.88     & 0.13              & 0.06                  & 44      & -77.47  & 58.6          & 37.0        & 1.00          & 0.08              & 0.76          \\
A1423 & 1.63     & 0.15              & 0.03                  & 66      & 63.53   & 2.9           & 6.8         & 0.85          & 0.03              & 0.87          \\
a209  & 1.63     & 0.27              & 0.03                  & -42     & -50.97  & 8.7           & 6.2         & 0.85          & 0.03              & 0.96          \\
A611  & 1.63     & 0.43              & 0.06                  & 38      & 36.53   & 1.8           & 33.5        & 0.95          & 0.12              & 0.95          \\
c1226     & 2.28     & 0.58              & 0.10                  & -79     & -85.89  & 6.5           & 11.7        & 0.31          & 0.17              & 0.71         \\
r2248     & 1.30     & 0.03              & 0.01                  & 64      & 50.34   & 13.3          & 7.2         & 0.94          & 0.01              & 0.95          
\enddata
\tablecomments{ This table includes the 20 X-ray selected CLASH clusters, ordered as in Table~\ref{table:phot}.}
\tablenotetext{a}{Position angles are reported with the convention of the longest axis pointing East of North, in degrees. The rotation solution provided in the text computes the angle of the first component (xx) from the x axis in the image. (For our mosaics, the x axis runs east-west.)}
\tablenotetext{b}{The columns labeled Centroid Diff and PA Diff contain the absolute magnitude of the differences between the IR and UV centroids and the IR and UV position angles, respectively,
as measured over identical regions on the sky. Note that while the UV and IR centroids are all within an arcsecond, only clusters with prominent star formation show significant ($>3\sigma$) differences in the position angles.}
\tablenotetext{c}{The axis ratio in the IR has negligible statistical uncertainty compared to the ratio in the UV. The ultraviolet emission appears to be somewhat more elongated than the IR for
star-forming galaxies; the outlier among the quiescent BCGs is in c1226 at $z=0.89$, which shows an interestingly elongated UV morphology in its core.}
\end{deluxetable*}

\section{Discussion}

\subsection{Star Formation Rates}

Table~\ref{table:phot} gives unobscured star formation rates based on UV photometry for CLASH BCGs. 
We have made an empirical estimate of the amount of UV
associated with the evolved stellar population, using the red UV--IR colors of the quiescent BCGs in our own sample. These colors are consistent with those found in previous work on low-redshift UV emission from BCGs 
\citep{2010ApJ...719.1844H,2010ApJ...715..881D,2012ApJS..199...23H}.
Our reported UV star formation rates have been adjusted for this contribution assuming a mean UV--NIR color for BCGs of 5.5. 
The uncertainty in this component translates to a  minimum systematic uncertainty in the star formation rate of $\sim 0.3$ \mdot for these galaxies. 

As we noted previously, the UV luminosity is a direct measurement of the unobscured star formation rate. 
But of course much of the star formation in a galaxy can be obscured. 
Galaxy studies indicate that for typical star-forming galaxies such as the Milky Way, 
the unobscured star formation rate is approximately equal to the obscured star formation rate, 
such that SFR$_{UV} \sim $ SFR$_{FIR}$. 
Typically, the unobscured (UV) star formation rate for a galaxy corresponds to 10--25\% of the total amount of star formation in galaxies with SFR $ \gtrsim 2 \, M_\odot \, {\rm yr}^{-1}$, based on comparisons with SFRs estimated 
from dust emission in the far infrared or H$\alpha$ emission \citep[e.g.,][]{2014MNRAS.441....2D,1998ARA&A..36..189K,2010ApJ...714.1256C}. 
While brightest cluster galaxies (BCGs) are either non-existent or very rare in these low-redshift multiwavelength studies, the existing data on BCGs indicate similar ratios of unobscured to total star formation \citep[e.g.,][]{2011ApJ...732...40D}. 

For the handful of BCGs with mid-infrared SFR estimates based on 
detections from Spitzer and Herschel \citep{2012ApJ...747...29R,2012ApJS..199...23H} 
the correlation between MIR SFR and UV SFR is strong \citep[e.g.,][]{2012ApJS..199...23H}. 
We note that several of the CLASH BCGs have nearly starburst-level luminosities (m1931, r1531, m0329) as indicated by  unabsorbed UV SFRs of $>25$ M$_\odot$ yr$^{-1}$. m1532 has a mid-IR based SFR $\sim 110$ \mdot\citep{2010ApJ...719.1619O, 2012ApJS..199...23H}, 
which is similar to but somewhat higher than the unabsorbed UV star formation rate of $\sim 80$ \mdot. 
Therefore the total (summed) star formation rates for CLASH BCGs could be 2--3 times higher than our estimates based on UV alone.

All of the BCGs with unobscured UV SFRs $ \gtrsim 1$ M$_\odot$ yr$^{-1}$ reside in 
clusters with low central gas entropy (entropy defined as $K=kT n_e^{-2/3}$).
We note that most of the central gas entropies reported in Table~\ref{table:phot} are from the 
ACCEPT sample \citep{2009ApJS..182...12C}, with the exception of m0416 and
m2129, which did not have public observations available at the time of the ACCEPT project. 
Those central gas entropies, which are large in both systems, are based on the 
X-ray gas profiles described in \citet{2014ApJ...794..136D}.

\begin{figure}
\includegraphics[width=3.5in, trim = 0.0in 0.0in 0.0in 0.0in,clip]{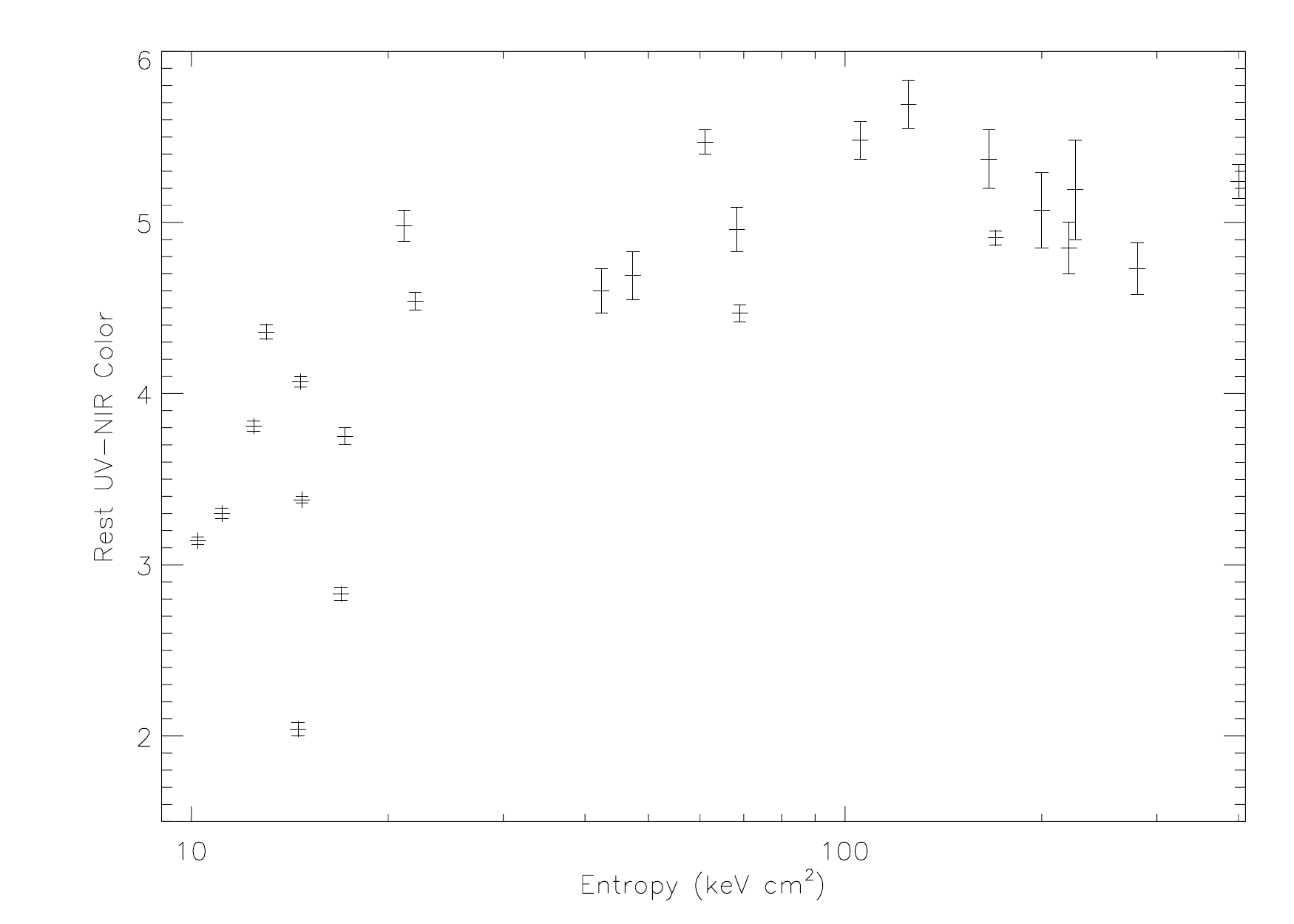}
\caption{\footnotesize 
Rest frame UV (280 nm)--NIR (1 micron) color of the central regions of CLASH BCGs plotted against central ICM gas entropy. (Gas entropy here is defined to be $K = T n_e^{-2/3}$, where $T$ is temperature in keV and $n_e$ is electron number per cubic cm.)
The connection between the thermodynamic state of circumgalactic gas surrounding the BCG and star formation
in the central galaxy is clear. The entropy threshold in the CLASH sample is similar to the entropy threshold seen
in lower redshift clusters ($z<0.2$) by Cavagnolo et al. (2009) and others.
\vspace*{1.0em}
\label{figure:UVSFR}}
\end{figure}

These findings for CLASH BCGs at $0.2 < z < 0.9$ are in accord with the well-known pattern observed among lower-redshift clusters. 
The fact that the thermodynamical state of the X-ray gas, and particularly the central cooling time, 
is strongly correlated with potentially star-forming optical emission-line gas was recognized 
as early as \citet{1985ApJS...59..447H}. 
Higher resolution X-ray imaging spectroscopy from Chandra allowed improved estimates of entropy profiles and central cooling times and led to the discovery of a well-defined threshold for multiphase gas and star formation at $K \sim 30$ keV cm$^2$ or $t_{cool} < 1 $ Gyr \citep{2008ApJ...683L.107C,2009ApJS..182...12C,Rafferty+08}. 
If cluster cores do not contain gas below these thresholds (which are nearly equivalent in clusters of galaxies with $kT \gtrsim $ a few keV), they do not host BCGs with strong radio sources or prominent star-formation signatures.
The fact that the same pattern holds among CLASH BCGs (Figure~\ref{figure:UVSFR}) shows that a similar entropy/cooling-time threshold has been in place since at least $z \sim 0.5$.

 \subsection{Morphologies \label{morph}}
 
The CLASH BCGs show a range of UV morphologies, and the nature of the morphology is correlated with UV luminosity and star-formation rate. 
BCGs with little to no star formation exhibit rather smooth UV morphologies similar to that of the underlying stellar population of evolved stars. 
In contrast, the UV morphologies of BCGs with UV excesses show clumps and filaments, indicating that active BCGs are forming stars, but not in disks or monolithic, single super-clusters. 
The clumps are likely to be clusters of hot stars, while the filaments could be unresolved strings of recently formed stars, or possibly emission-line filaments corresponding to interfaces between $10^7$~K and $10^4$~K gas. 
These structures are similar to those seen in the rest-frame UV images of BCGs \citep{2010ApJ...719.1619O} and in ground-based blue images \citep{1997ASPC..115..109M}. \citet{1997ASPC..115..109M} identify four structural features: point sources, lobes, disks, and amorphous sources. As found in \citet{2010ApJ...719.1619O} and in \citet{1997ASPC..115..109M}, most if not all of the star-forming systems in our CLASH sample would fall into the amorphous category (with multiple clumps). Only a383 and m1720 have linear structures that might be very small disks. 

Figure \ref{UVimage1} shows UV images of 9 of the 11 systems with core entropy $< 30 \, {\rm keV \, cm^2}$ (excluding ms2137 and r2129), in order of unobscured star-formation rate.  All of them show significant structure, and the amount of structure increases with increasing star-formation rate.  The most extreme star-forming BCGs exhibit the largest UV extents and biggest centroid shifts in comparisons among UV and near-IR morphological quantities. 
The other star-forming BCGs are somewhat more compact than the most UV-luminous BCGs, and are in general fairly well centered on the near-IR images. 
In these less vigorously star-forming BCGs, the UV emission is typically less extended (smaller-scale moments) than the near-IR light, even within a limited aperture centered on the BCG. 
The higher entropy systems in Figure~\ref{UVimage2} are more uniform in appearance, with smooth, symmetric, centrally concentrated UV emission profiles.  UV emission in these systems tracks the the evolved stellar population, which dominates the near-IR light emitted by both star-forming and quiescent BCGs.

Table~\ref{table:morph} reports morphological parameters that quantify these differences between quiescent and star-forming BCGs. 
UV morphological parameters of quiescent BCGs are very similar to those measured in 
the near-IR images, while star-forming BCGs exhibit different axis ratios and sometimes different
centroids in the UV compared to the near-IR.   
Position angles derived from the infrared and the UV images are mostly well correlated. 
Large position-angle differences found in a few of the quiescent BCGs are not significant because those galaxies are nearly round, leading to large uncertainties in the position angles.  
However, 5 of the 6 BCGs with unobscured star-formation rates $> 10$~\mdot have UV position angles that differ from the IR position angles by more than 3$\sigma$.

\begin{figure}
\includegraphics[width=3.5in, trim = 0.0in 0.0in 0.0in 0.0in,clip=true]{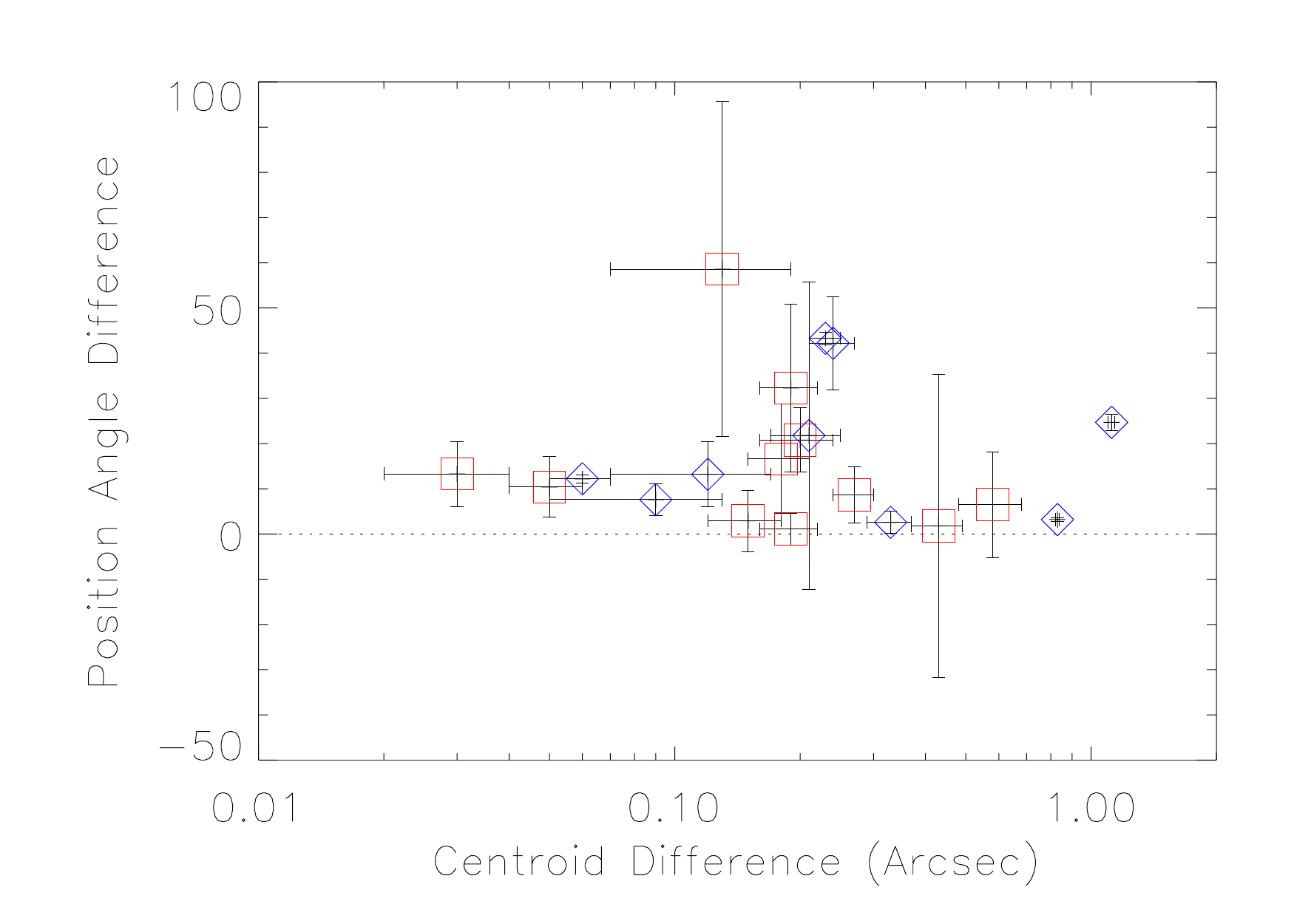}
\caption{\footnotesize
Centroid and position-angle differences between UV and near-IR emission for the 20 X-ray selected CLASH BCGs.  Blue diamonds show star-forming BCGs.  Red squares show quiescent BCGs.  Error bars show 1-sigma uncertainties.
Position-angle differences in position angles are usually consistent with zero, given the uncertainties, except for the star-forming BCGs with highly structured UV morphologies.   
Also, the objects with the largest and most significant centroid differences between UV and near-IR emission are star-forming BCGs. 
\label{figure:ScaleRatio}}
\end{figure}

\subsubsection{Morphological Details of Star-Forming CLASH BCGs}

The most extreme SF BCGs in CLASH are in clusters m1931 and r1532. 
m1931 has multiple clumps, with the largest and most luminous clumps in the center. There are several clear
filaments, aligned preferentially north/south. The knot detail is also visible in continuum images, indicating that
the stellar density is also enhanced in those regions, not just the emission lines. The UV structures extend over 20 kpc 
NW of the BCG, and somewhat to the SE, for an overall elongated appearance. 
There is a curved, clearly resolved dust lane
cupping the southeast quadrant of the clumpy nuclear region, while the brightest three non-nuclear clumps spray
to the NE, N, and NW. This BCG exhibits GALEX emission at even shorter wavelengths 
(Table~\ref{table:GALEX_NVSS}). R1532 also has radial filaments 
extending $\gtrsim 20$ kpc in all directions, but the longest filaments extend to the south, 
and multiple clumps surround the nucleus and extend southward.
Among the CLASH BCGs, this one probably exhibits the most similarity to NGC1275. The brightest
of its knots has been targeted for HST COS photometry and STIS long-slit observations later this year (PI: Donahue.) This BCG also exhibits curved, comma-shaped dust lanes wrapping the central knots; dust lanes appear alongside filaments and knots.

M0329 exhibits several clumps and a prominent curved filament extending north from a 
flattened structure. It lacks FIR observations but
is likely a luminous IR source, given that its unobscured star-formation rate is similar to those of other BCG LIRGs.
It was also a GALEX detection, with a 7.5 arcsecond diameter AB mag of $21.75 \pm 0.22$ (NUV) and $22.6 \pm 0.6$ (FUV). 

M0429 shows what may be two distinct, clearly separate, clumps interacting in the center of cluster.
Several other BCGs have a compact, multi-clump morphology. M1115 has 3-4 clumps oriented like beads
on a curved string, with a couple of fainter sources a couple arcseconds off-center. M1423 has a bright
but extended central source sitting in an elliptical distribution of fainter UV light.  It
is also a luminous UV source in the observer frame, with rest-frame far UV detections of $22.0\pm0.2$ and
$23.1 \pm 0.6$ AB mag (Table~\ref{table:GALEX_NVSS}). Note that for a BCG at $z=0.545$, the NUV and FUV bandpasses of GALEX
cover mean rest-frame wavelengths of 147 and 98 nm respectively. 
In the F275W image, rest-frame far-UV ($\lambda \sim 180$ nm), which is the bluest image CLASH obtained for this cluster, 
there is an elongated (0.3-0.5 arcsecond) nuclear object located precisely with the BCG, oriented about 35 degrees W of N. 
Its IR estimated SFR rate of $46 \pm 3$ \mdot \citep{2012ApJ...747...29R} 
is 3 times the unobscured UV rate in Table 2, an IR excess typical of obscured starburst galaxies.

R1347 is a very well-known active BCG in what was once the most luminous X-ray cluster known
\citep{1995A&A...299L...9S} and has a cool core with evidence for a merger \citep{2010ApJ...716..739M}. 
It does not show a spectacular UV filament system, at least not as
spectacular as other CLASH BCGs in this category. But it does have a very bright SF 
region just to the ESE of the nucleus, and a faint filament on the opposite side of the
nucleus.

\subsection{Comparison with Simulations of AGN Feedback \label{agn}}

Cooling and heating of the circumgalactic medium (CGM) around BCGs has been a topic of
much interest for 30--40 years \citep{1994ARA&A..32..277F}. AGN feedback is increasingly thought to be the key physical process regulating the CGM \citep{2007ARA&A..45..117M,2012NJPh...14e5023M},  and the CLASH clusters with low-entropy cores all appear to have BCGs with active nuclei.
 In the CLASH sample, 10 of the 11 clusters with $K_0 < 30$ keV cm$^2$ have potential AGN counterparts in the 1.4 GHz NRAO VLA Sky Survey  \citep[hereafter NVSS][]{1998AJ....115.1693C} within 7 arcseconds of the BCG.  
The only low-$K_0$ cluster lacking an NVSS counterpart (m1931) has a luminous X-ray and radio 
AGN in its core \citep{2011MNRAS.411.1641E, 2014ApJ...794..136D}. (The m1931 BCG radio source is probably missing from the NVSS because of blending with the more luminous Narrow-Angle Tailed (NAT) radio galaxy about $30\arcsec$ south of the BCG. See the VLA radio map in \citet{2011MNRAS.411.1641E}.)
In contrast, only 3 of the other 13 clusters that are members of both the CLASH sample and
the NVSS sky coverage have NVSS radio sources with centroids as close to the BCG. 
One of these clusters is m1206, which hosts a known lensed feature in the core of its BCG, so future high-resolution radio observations and optical spectroscopy are required to distinguish the lensed component from 
intrinsic radio emission. 
The other two are c1226, which has three clumps in the UV image of its BCG, and A2261, which shows 
evidence for ``scouring" by the merging of supermassive black holes in the BCG core \citep{2012ApJ...756..159P}.
The only CLASH cluster too far south to be in NVSS's sky coverage, r2248, also lacks a counterpart within an arc minute of the BCG in the 843 MHz Sydney University Molongolo Sky Survey 
\citep[][known as SUMMS]{1999AJ....117.1578B,2003MNRAS.342.1117M}.  
It also has only upper limits for its far-IR or submm flux from Herschel \citep{2012ApJ...747...29R}. 
Also, the two most extreme star-forming BCGs in CLASH show evidence in deep Chandra observations for X-ray cavities near the central radio source \citep{2011MNRAS.411.1641E,2013ApJ...777..163H}. 

Recent hydrodynamical simulations of precipitation-regulated feedback in BCGs  \citet{2014ApJ...789...54L,2014ApJ...789..153L} produce star-forming structures of cold gas with extents and morphologies very similar to the structures we observed in UV images of star-forming CLASH BCGs.  
In these simulations, thermal instabilities produce cold clouds that rain into the central AGN, which responds by propelling  bi-polar jets into the ambient medium. 
Interestingly, the circumgalactic medium initially responds to a jet outburst by producing additional cold gas, as the jets uplift low-entropy gas from the central region of the BCG to higher altitudes and cause it to condense into cold, star-forming gas clouds. 
Eventually the cold gas relaxes into a disk, but after that disk of gas is consumed by star formation, the AGN reignites 
as the cooling cycle re-establishes itself \citep{2015arXiv150302660L}. 
In these simulations, gas flows in and out of the central region of the galaxy, and the stars preferentially form along
the jet axis with a radial extent similar to what we see in the UV images of CLASH BCGs.

Figure~\ref{UVimage1} shows a montage of simulated temperature maps beneath the montage of UV images of CLASH SF BCGs. 
The coldest gas, which shows up as blue in these maps, indicates where stars would be forming in the simulation. 
We have selected frames from a movie of a single simulated cluster, in which the time indicator in the upper left corner gives the time since the beginning of the simulation. 
The first major AGN outburst in this simulation happens $\sim300$ Myr after the beginning. 
Frames were selected to be similar in appearance to the CLASH UV images in the upper part of the figure. 
Our intention is to illustrate the various morphologies that can emerge during a single episode of AGN feedback and to show how similar they can be to the UV morphologies of star-forming BCGs. 
A movie of this simulation can be viewed at \url{https://vimeo.com/84807876}. 
 A similar picture of circulating, heating and cooling gas, moderated by the influence of X-ray cavities in the hot atmosphere of groups and clusters is discussed in \citet{2015arXiv150107647B}.
 
Interestingly, \citet{2005AJ....129.2138L} infer similar timescales and outflow phenomena from the frequency and morphology of dust structures revealed by HST optical imaging of their sample of 77 elliptical galaxies. The dust structures seen in imaging at similar wavelengths in CLASH BCGs are not identical to the UV structures, but show similar extents. We will provide further details on the dust morphology and resolved structures in an upcoming paper and PhD thesis by Kevin Fogarty.

\section{Conclusions}

The uniquely wide bandpass coverage and spatial resolution of the CLASH HST dataset covering 25 clusters with $z \sim 0.2-0.9$ 
provides a unique view into the UV appearance of brightest cluster galaxies in some of  the most massive clusters
in the universe. 
Contrary to the quiesescent, red-and-dead reputation of BCGs, their UV appearances exhibit a wide diversity of morphologies, similar to those found in the small sample of extreme cool-core BCGs that have been studied in the rest-frame UV at low redshifts. 
The only CLASH clusters to host BCGs with detectable unobscured star formation are those with
low-entropy X-ray emitting gas in their centers: the ``cool-core" clusters. 
The BCGs with large star-formation rates are also the only ones with knots, filaments, and other asymmetric features in their UV morphologies. 
One exception is m1206, with a central feature that is likely to be related to a strongly lensed galaxy that is also responsible for narrow, radial filaments in optical images.

UV emission from quiescent BCGs is less bright but still detectable, showing a smooth, centrally symmetric distribution with a typical UV$_{280}$ - NIR color of $\sim 5-5.5$, which is consistent with low-redshift studies using GALEX and the XMM Optical Monitor.
Some of these galaxies show moderate UV excesses, which are probably due to the small but finite intrinsic scatter in the UV-NIR color of the older, evolved population of stars.
The unobscured star-formation rates we estimate, after correcting for this UV component from the elderly population, may be underestimates of the total star formation rates by a factor of 2 or more.  If this underestimate increases with increasing star-formation rate, as in other strongly star-forming galaxies, the star-formation rates in the most extreme CLASH BCGs likely exceed 100 \mdot.

The entropy or cooling-time threshold for star formation in the CLASH BCGs is similar to that seen in low-redshift ($z<0.2$) clusters by Cavagnolo and others. 
 While we cannot rule out the possibility of a somewhat lower entropy threshold $\sim 20$~keV~cm$^2$, it is interesting to note that none of the high-entropy clusters in our sample have BCGs with excess UV light or clumpy UV morphologies, suggesting that even though our X-ray data are limited, we are not missing significant low-entropy cores in the CLASH clusters. 
Two potential exceptions worth following up in detail are c1226 at $z=0.89$, which is extremely hot and has a BCG showing some indication of multiple components along the east-west axis, and m1206, in order to sort out the lensed contribution of its very interesting central UV/radio source.

The high resolution UV rest-frame images of star-forming BCGs show a delightful variety of morphologies and structures.  The scales and structures of star-forming regions seen in these UV images are similar to those of cold-gas structures in recent simulations of AGN feedback in idealized BCGs, suggesting that while mergers and interactions
may play a role in the appearance of these star-forming regions,  AGN feedback is a contributing if not a dominant driver of  the morphology and distribution of star formation in BCGs.

\acknowledgments

We acknowledge Adi Zitrin for helpful comments on the text.
Support for the CLASH MCT program (ID 12100) was provided by NASA through grants from the Space Telescope 
Science Institute, which is operated by the Association of Universities for Research in 
Astronomy, Inc., under NASA contract NAS 5-26555. YL acknowledges financial support from 
NSF grants AST-0908390, AST-1008134, AST-1210890 and NASA grant NNX12AH41G,  as well 
as computational resources from NASA, NSF XSEDE and Columbia University.

\begin{appendix}

\subsection{Lensed UV Features Coinciding with the BCG}

Two of the BCGs show UV features that are lensed background galaxies, unrelated to
the cluster: m1206 and m1149. We discuss these here because they are very prominent
UV features in the HST imagery, but likely have very little to do with the BCG itself.

The BCG in m1206 shows two extremely long narrow filamentary features,
one curved and one nearly radial, diving into the nuclear region. The center of the BCG 
has a curled tadpole-type morphology, which is still visible even in the bluest UV images
available.  As a relatively high central gas entropy system of  $K_0 \sim 70$ keV cm$^2$, 
it would have been surprising if its BCG turned out to host new stars. Two sets of authors
have identified these features as likely to be strongly lensed background galaxies with 
uncertain photometric redshifts of around $z\sim1.2-1.7$ \citep{2012ApJ...749...97Z,2013ApJ...774..124E}.

The BCG in the cluster core of m1149 is blended with 
a beautifully lensed, $z=1.491$ background spiral galaxy, 
identified in archival data by \citet{2009ApJ...703L.132Z} and confirmed spectroscopically 
by \citet{2009ApJ...707L.163S}. The  Lyman $\alpha$ emission 
from this galaxy falls into the rest-frame UV of the BCG. 

\subsection{BCG Notes and Relevant Survey Data}

 The BCG of m0717 is non-trivial to identify unambiguously since it is neither dominant nor centrally located.  The candidate we selected 
 is one of the brightest galaxies spectroscopically identified to be a member
 of the cluster by \citet{2014ApJS..211...21E}, at an RA (J2000) of 07:17:32.518 and Dec (J2000) of +37:44:34.84,
 with $z_{spec}=0.5415$. This galaxy is located away the nominal center of the red galaxies from the CLASH list in \citet{2012ApJS..199...25P}
 and \citet{2013ApJ...777...43M}. There are no BCG candidates near this position. The galaxy identified as the BCG for this cluster by
 \citet{2012ApJ...747...29R} (associated with a Herschel mid-IR upper limit) is likely to be a foreground elliptical galaxy.
 For completeness, we list the central locations of all of the BCGs utilized for UV and near IR photometry in Table~\ref{table:GALEX_NVSS}. Most
 of these coordinates are nearly identical to those in Postman et al. (2012).
 
We cross-matched the CLASH BCG locations (in Table~\ref{table:GALEX_NVSS}) with the NVSS \citep[][]{1998AJ....115.1693C} and 
SUMSS \citep[][]{1999AJ....117.1578B,2003MNRAS.342.1117M} catalogs of radio sources, with the criterion that a source should
be aligned within $7\arcsec$ to be considered a candidate counterpart. We estimated the rest-frame 1.4 GHz luminosities (W Hz$^{-1} h_{70}^{-2}$) by assuming
that $L_\nu \propto \nu^{-1}$, such that $L_{1.4\rm{GHz}} = 4 \pi D^2 (1+z)^2 F_{1.4\rm{GHz}}$, where $D = \mathcal{R} \sin{r/\mathcal{R}}$ is the distance measure to the source, in which $r$ is the comoving radial distance coordinate and $\mathcal{R} \gg r$ is the cosmological radius of curvature. 
The beams of both of these surveys are quite large, $45\arcsec$. In one case, m1931, we found that
the BCG source was outshined by a NAT source in the field, and so the NVSS does not record a radio source near m1931, when in 
fact VLA observations showed a 70 mJy source there. We completed this search to show that all of the BCGs with UV excesses and
irregular UV morphologies as seen by HST have AGN activity from a relatively powerful central radio source, and we present the results
in this appendix.

The GALEX \citep{2007ApJS..173..682M} archive increased in size since the ACCEPT-based work on BCGs by Hoffer et al. (2012). Therefore, 
we also cross-matched the BCG locations with the final (GR6/7) GALEX source list, utilizing the GALEXview service\footnote{\url{http://galex.stsci.edu/GalexView/}}
from MAST\footnote{The Barbara Mikulski Archive for Space Telescopes, \url{http://archive.stsci.edu}}, and report the results in Table~\ref{table:GALEX_NVSS}. The fluxes
are estimates of the total flux of the source in AB magnitudes in the NUV bandpass (227 nm, 62 nm FWHM) and in the FUV (152 nm, 27 nm FWHM). 
The angular resolution of GALEX (about $5\arcsec$) is considerably worse than that of HST, so
very little morphological information can be gleaned for GALEX for our targets, and as with the radio, there is a danger of spurious associations owing to confusion 
between source and backgrounds/foregrounds. 
We note that since GALEX was an ultraviolet telescope, the central wavelength of its NUV
bandpass for CLASH clusters sits almost 100 nm farther to the ultraviolet. For about 1/2 of the sample, Lyman-$\alpha$ is included
in the NUV bandpass of GALEX so any SFRs based on these numbers
would be quite uncertain. (Since the blue continuum of stars is approximately flat in $L_\nu$ to 150 nm, a similar AB magnitude for continuum
measured at 170 nm would correspond to a similar star formation rate measured at 280 nm \citep{1998ARA&A..36..189K}, modulo an intrinsic reddening uncertainty.)
Nevertheless we note that given the very different effective apertures and wavelengths, there is consistency with the upper limits and detections by GALEX with 
the rest-frame UV 280 nmphotometry we have obtained from the HST data. We applied approximate
Galactic reddening corrections, following our procedure in Hoffer et al. (2012); we report the fluxes corrected for Galactic
reddening in Table~\ref{table:GALEX_NVSS}. Three-sigma upper limits for exposures up to 500
seconds were estimated from the relations estimated Hoffer et al. (2012). Upper limits for BCGs lacking cataloged detections in 
longer GALEX exposures were estimated based on the fluxes of the faintest detected GALEX catalog sources within 
10 arcminutes of the source with flux uncertainties $<0.35$ AB mag ($\sim 3\sigma$).

\begin{deluxetable*}{lllllllllllll}
\tablecaption{GALEX EUV and NVSS Radio Fluxes \label{table:GALEX_NVSS}}
\tablehead{
\colhead{Name}        & \colhead{BCG} & \colhead{BCG} & \colhead{NUV} & \colhead{FUV} &  \colhead{E(B-V)}  & \colhead{A$_{NUV}$} & \colhead{A$_{FUV}$} & \colhead{NUV\tablenotemark{c}}  & \colhead{FUV\tablenotemark{c}} 
					& \colhead{NVSS} & \colhead{offset} & \colhead{Log L$_{1.4\rm{GHz}}$}    \\
\colhead{    }        & \colhead{RA(J2000)} & \colhead{DEC(J2000)} & \colhead{AB} & \colhead{AB} & \colhead{E(B-V)}  & \colhead{mag} & \colhead{mag} & \colhead{AB}  & \colhead{AB} 
					& \colhead{mJy} & \colhead{($\arcsec$)} & \colhead{W Hz$^{-1}$}    } 
\startdata 
a209   & 01:31:52.54 & -13:36:40.4 & $21.14          \pm 0.08$ & $22.04         \pm 0.07$      & 0.0109 & 0.11   & 0.08      & 21.25     & 22.12   & \textless2.5                &  & \textless23.5           \\
A383   & 02:48:03.40 & -03:31:44.9 & $19.8           \pm 0.1$  & $18.72         \pm 0.12$      & 0.0315 & 0.32   & 0.24      & 20.12     & 18.96   & 40.9                        & 1         & 24.6           \\
m0329 & 03:29:41.56 & -02:11:46.4 & $20.96          \pm 0.08$ & $21.25         \pm 0.12$      & 0.0615 & 0.62   & 0.48      & 21.58     & 21.73   & 6.9                         & 6.67      & 24.7           \\
m0416 & 04:16:09.13 & -24:04:03.0 & $>22.1$        & $>21.4$ &                            0.0413 & 0.42   & 0.32      & $>22.5$      & $>21.7$    & \textless2.5                &  & \textless24.1           \\
m0429 & 04:29:36.05 & -02:53:06.1 & \nodata              & \nodata       &               &  \nodata   &  \nodata      & \nodata   & \nodata & 138.8                       & 0.71      & 25.9           \\
m0647 & 06:47:50.64 & +70:14:54.0 & $>21.6$        & $>22.0$ &                            0.1103 & 1.11   & 0.85      & $>22.7$      & $>22.9$    & \textless2.5                &  & \textless24.5           \\
m0717 & 07:17:32.52 & +37:44:34.9 & $>22.1$        & $>22.5$ &                            0.0582 & 0.59   & 0.45      & $>22.7$      & $>23.0$    & \textless2.5                &  & \textless24.5           \\
m0744 & 07:44:52.82 & +39:27:26.9 & $>22.0$        & $>24.3$ &                            0.0578 & 0.58   & 0.45      & $>22.6$      & $>24.7$    & \textless2.5                &  & \textless24.7           \\
A611   & 08:00:56.82 & +36:03:23.6 & $>23.0$        & $>23.1$ &                            0.0573 & 0.58   & 0.44      & $>23.6$      & $>23.5$    & \textless2.5                &  & \textless23.8           \\
m1115 & 11:15:51.90 & +01:29:55.4 & $21.17          \pm 0.08$ & $21.6          \pm 0.1$       & 0.0386 & 0.39   & 0.30      & 21.56     & 21.90   & 16.2                        & 1.47      & 24.8           \\
m1149 & 11:49:35.69 & +22:23:54.6 & $>21.5$        & $>21.0$ &                            0.0229 & 0.23   & 0.18      & $>21.7$      & $>21.2$    & \textless2.5\tablenotemark{b}                &  & \textless24.5           \\
A1423  & 11:57:17.36 & +33:36:39.6 & $>21.6$        & $>21.1$ &                            0.0192 & 0.19   & 0.15      & $>21.8$      & $>21.2$    & \textless2.5                &       & \textless23.5           \\
m1206 & 12:06:12.09 & -08:48:03.3 & $>21.5$        & $>21.0$ &                            0.065  & 0.65   & 0.50      & $>22.2$      & $>21.5$    & 160.9                       & 1.94      & 26.1           \\
c1226 & 12:26:58.25 & +33:32:48.6 & $23.9           \pm 0.17$ & $>25$                              & 0.0187 & 0.19   & 0.14      & 24.09     & $>25.1$    & 4.3                         & 4.37      & 25.2           \\
m1311 & 13:11:01.80 & -03:10:39.8 & $>22.4$        & $>23$   &                            0.0304 & 0.31   & 0.24      & $>22.7$      & $>23.2$    & \textless2.5                &  & \textless24.4           \\
r1347 & 13:47:30.62 & -11:45:09.4 & $21.57          \pm 0.07$ & $21.82        \pm 0.08$      & 0.0615 & 0.62   & 0.48      & 22.19     & 22.30   & 45.9                        & 1.09      & 25.5           \\
m1423 & 14:23:47.88 & +24:04:42.5 & $21.3           \pm 0.3$  & $>21.4$            & 0.0277 & 0.28   & 0.21      & 21.58     & $>21.6$    & 8                           & 2.59      & 25.0           \\
m1532 & 15:32:53.78 & +30:20:59.4 & \nodata              & \nodata                                  &  \nodata      &  \nodata   &  \nodata      & \nodata   & \nodata & 22.8                        & 0.48      & 25.0           \\
m1720 & 17:20:16.78 & +35:36:26.5 & $>21.8$       & $>21.2$           &                  0.0374 & 0.38   & 0.29      & $>22.2$      & $>21.5$    & 18                          & 1.93      & 25.0           \\
A2261   & 17:22:27.18 & +32:07:57.3 & \nodata              & \nodata                                  &   \nodata      &  \nodata   &  \nodata      & \nodata   & \nodata & 5.3                         & 6.15      & 23.9           \\
m1931 & 19:31:49.65 & -26:34:32.8 & $19.49          \pm 0.11$ & $19.2          \pm 0.1$       & 0.1107 & 1.12   & 0.86      & 20.61     & 20.06   & 70                          &  & \textless24.0           \\
m2129 & 21:29:26.11 & -07:41:27.7 & $>22.1$        & $>21.4$           &                  0.0759 & 0.76   & 0.59      & $>22.9$      & $>22.0$    & \textless2.5                &  & \textless24.5           \\
r2129 & 21:29:39.96 & +00:05:21.2 & $21.85          \pm 0.12$ & $21.28         \pm 0.07$      & 0.0402 & 0.41   & 0.31      & 22.26     & 21.59   & 25.4                        & 1.76      & 24.6           \\
ms2137  & 21:40:15.17 & -23:39:40.2 & $22.09          \pm 0.05$ & $21.89         \pm 0.03$      & 0.0511 & 0.51   & 0.40      & 22.60     & 22.29   & 3.8                         & 2.52      & 24.1           \\
r2248 & 22:48:43.96 & -44:31:51.1 & $>21.90 $      & $>21.4$                            & 0.0122 & 0.12   & 0.09      & $>22.0$      & $>21.5 $   & \textless6\tablenotemark{a} &  & \textless24.2          
\enddata
\tablenotetext{a}{SUMMS 843 MHz limit. All the others are from the NVSS at 1.4 GHz.}
\tablenotetext{b}{The BCG in m1149 may have a nearby radio counterpart, offset about $10\arcsec$ in the sky. M1149 is a merging cluster, so the radio source may be related to merger processes or the BCG; it could also be unrelated.}
\tablenotetext{c}{GALEX NUV and FUV total fluxes and upper limits reported in these columns are corrected for Galactic extinction.}
\end{deluxetable*}

\end{appendix}

{\it Facilities:} \facility{Hubble, Chandra}

\bibliography{CLASH_UV}

\end{document}